\documentclass[showpacs,preprintnumbers,amsmath,amssymb,superscriptaddress,nofootinbib,english]{revtex4}
\usepackage{graphicx}
\usepackage{dcolumn}
\usepackage{bm}
\usepackage{epsfig}
\usepackage[usenames]{color}
\usepackage{url}

\newcommand{\remove}[1]{}

\def\be{\begin{equation}}
\def\ee{\end{equation}}
\def\ba{\begin{eqnarray}}
\def\ea{\end{eqnarray}}

\begin{document}

\title{ Brane Isotropisation in  Extra-Dimensional Tolman-Bondi Universe}

\author{Philippe~Brax}
\email[]{philippe.brax@cea.fr}
\affiliation{Institut de Physique Theorique, CEA, IPhT, CNRS, URA 2306, F-91191Gif/Yvette Cedex, France}

\author{Jos\'e P. Mimoso}
\email[]{jpmimoso@cii.fc.ul.pt}
\affiliation{Faculdade de Ci\^encias e Centro de Astronomia e Astrof\'isica da Universidade de Lisboa, 1749-016 Lisboa, Portugal}

\author{Nelson J. Nunes}
\email[]{njnunes@fc.ul.pt}
\affiliation{Faculdade de Ci\^encias e Centro de Astronomia e Astrof\'isica da Universidade de Lisboa, 1749-016 Lisboa, Portugal}

\date{\today}

\begin{abstract}
We consider the dynamics of a 3-brane embedded in an extra-dimensional Tolman-Bondi Universe where the origin of space plays a special r\^ole. The embedding is chosen such that the induced matter distribution on the brane respects the  spherical symmetry of matter in the extra dimensional space. The mirage cosmology on the probe brane  is
studied, resulting in an inhomogeneous and anisotropic four dimensional cosmology where the origin of space is also special. We then focus on the spatial geometry around the origin and show that the induced geometry, which is initially inhomogeneous and anisotropic,  converges to an isotropic and homogeneous Friedmann-Lemaitre 4d space-time. For instance, when a 3-brane is embedded in a 5d matter dominated model, the 4d dynamics around the origin converge to a Friedmann-Lemaitre Universe in a radiation dominated epoch.
We analyse this  isotropisation process and show that it is a late time attractor.
\end{abstract}

\maketitle

\section{Introduction}

The possible existence of extra dimensions is an exciting prospect. Ever since the first unification attempt by Kaluza \cite{Kaluza:1921tu} and Klein \cite{Klein:1926tv}, and then the advent of string theory, the role of these extra dimensions in shaping four dimensional physics has been emphasized \cite{Green:1987sp,Forste:2001ah,Aharony:1999ti}. It could range from the presence of towers of excitations in the particle spectrum to the appearance of scalar fields in the form of compactification moduli. In the last fifteen years and in particular after the discovery of fundamental branes and the building of phenomenological brane-world models \cite{Brax:2003fv} such as the one constructed by Randall and Sundrum \cite{Randall:1999ee,Randall:1999vf}, the emerging possibility of extra dimensional cosmology has been considered thoroughly. In brane cosmology, a brane configuration comprising one or several branes are embedded in a fixed extra dimensional background. This extra dimensional background is commonly assumed to be either a Minkowski space-time or an anti de Sitter Universe (for another type of background see \cite{Brax:2002qw}  for instance). These two cases receive strong motivation from either supergravity or string theory in more than four dimensions. When the brane is not taken as a probe brane, the extra dimensional geometry can be influenced by the matter fields on the brane and a back-reaction must be taken into account. In five dimensions, these effects are taken into account by imposing the appropriate Israel boundary condition \cite{Brax:2003fv}. On the other hand, for branes with a low enough tension, and/or no matter on their world surface, the branes whose dynamics are governed by their Dirac-Born-Infeld action evolve in the extra dimensional background leading to a mirage cosmology on the brane \cite{Kehagias:1999vr,Alishahiha:2004eh}. In particular, the cosmology on the brane defined by the induced metric can be non-trivial, e.g. inflation can result for instance.

Recently, the discovery of the acceleration of the Universe
\cite{Perlmutter:1998np} has led to a reappraisal of the underlying assumptions of cosmology. In particular, the cosmological principle, which states that the Universe is homogeneous and isotropic on large scales, has been under scrutiny \cite{Brax:2009ae}. As a result, Tolman-Bondi models of the Universe where the earth would lie close to the centre of an inhomogeneous Universe have been considered
\cite{Zehavi:1998gz,Tomita:2000jj,Caldwell:2007yu}. Along these lines, it seems to be timely to question the usual hypothesis that extra dimensions should be a symmetric space. In 5d, homogeneous and anisotropic brane models have been extensively studied \cite{Fabbri:2004mi,ver,campos,fro}, in particular with the aim of extending Wald's theorem which states that a space-time with a positive cosmological constant and a matter density satisfying the strong energy condition always isotropises \cite{wald}. Other approaches such as in  \cite{Watson:2002nx} tackle the issue of isotropisation of three large dimensions  in the brane gas cosmology  framework by allowing an arbitrary amount of initial anisotropy. There it  was  found that  the anisotropy  reaches a maximum early in the evolution and then approaches zero at later times.

Here and in the following, we will consider that the Universe is a 3-brane embedded in an extra dimensional space with no homogeneity at all. We study the dynamics of the brane as it responds to the extra-dimensional cosmology. To simplify matters, we assume that the $(d+4)$ dimensional space-time is a matter dominated, inhomogeneous and isotropic Tolman Bondi Universe with a special role played by the origin of space. On the brane, the induced metric is both inhomogeneous and anisotropic. In principle, we should also introduce matter fields on the brane and study the back reaction problem. As a first step, we will assume that the brane is a probe and as a result we only investigate the mirage cosmology on the brane. It turns out that the dynamics are governed by non-linear partial differential equations whose long time behaviour can be analysed in the vicinity of the origin. We find that the asymptotic behaviour of the induced metric around the origin becomes isotropic and homogeneous. This isotropisation on the brane is in fact an attractor behaviour.

Of course our model does not directly apply to our 4d Universe. Strong constraints on isotropy during Big Bang Nucleosynthesis should be taken into account \cite{Kolb:1985sj,Li:2005aia}. Moreover, the dynamics would be certainly changed by the presence of matter on the brane. Nevertheless, the fact that an isotropic and homogeneous space-time emerging from an initially inhomogeneous and anisotropic Universe around the origin of space is certainly appealing.

The outline of the paper is as follows. In a first part, we describe branes in a Tolman Bondi space-time. In section III, we study the mirage cosmology of a neutral brane in a Tolman Bondi Universe. In section IV, we simplify the dynamics and focus on the vicinity of the origin of space. In this case, the resulting inhomogeneous and anisotropic cosmology converges to a 4d Friedmann-Lemaitre Universe. Our results are confirmed numerically.

\section{Branes in Tolman-Bondi}

\subsection{Brane Dynamics}

We consider the dynamics of a single uncharged 3-brane described by its world-volume action
\begin{equation}
\label{3braneaction}
S_b=-T_3 \int d^4x \sqrt{-\tilde  g} ,
\end{equation}
where $T_3$ is the brane tension and $\tilde g_{\mu\nu}$ is the induced metric on the brane related to the extra dimensional metric $G_{AB}$ by
\be
\label{inducedmetric}
\tilde g_{\mu\nu}= G_{AB}\frac{\partial X^A}{\partial x^\mu}\frac{\partial X^B}{\partial x^\nu}
\ee
where $x^\mu$ are the brane coordinates and $X^A(x^\mu)$ the brane embedding.
Previous studies have mainly concentrated on the embedding of
3-branes in Minkowski  or anti de Sitter space-times. Here we concentrate on the possibility that matter may exist in the extra-dimensions and may lead to a dynamical time evolution of the extra-dimensional geometry. In the extra-dimensions (the bulk), the action is simply
\begin{equation}
S_{4+d}= \frac{1}{2\kappa_{d+4}^2}  \int d^{d+4}X \sqrt{-G} R_{d+4} + S_{d+4}^{\rm matter}.
\end{equation} Notice the dimension of $\kappa_{d+4}^2$ is $-(d+2)$. The matter action may comprise ordinary matter like a pressure-less fluid.
We will also assume that the distribution of matter in the bulk could be inhomogeneous, i.e. the cosmological principle could be violated in the bulk. If this is the case, then the brane geometry will be both anisotropic and inhomogeneous. Even if the cosmological principle is satisfied in the bulk, we will find that the brane geometry is generically anisotropic.

In order to retrieve gravity in the 4d sense at least in the low energy-long distance regime in a way akin to the Dvali-Gabadadze-Porrati (DGP) model \cite{Dvali:2000hr}, one may include an Einstein-Hilbert term on the brane and couple 4d gravity to matter on the brane
\begin{equation}
S_4=  \frac{1}{2\kappa_{4}^2} \int d^{4}x \sqrt{-\tilde g} \tilde R_{4} + S_{4}^{\rm matter},
\end{equation}
where $\tilde R_{4}$ is the Ricci scalar of the induced metric on the brane and $\kappa_4$ the 4d gravitational constant.
Here $S_{4}^{\rm matter}$ is the contribution from the matter species living on the brane.

In the following we will treat the simpler case where the effects of brane gravity and matter are neglected. In this mirage cosmology setting where the dynamics on the 3-brane are simply due to its probe nature and its motion in the bulk, we will be able to analyse interesting phenomena such as the isotropisation of the 4-d geometry.
Of course, the fact that we do not take into account $S_4$ means that our model cannot describe the physics of the Universe since Big Bang Nucleosynthesis. To do so, a full analysis of the model involving the total action
\begin{equation}
S_T= S_{d+4} + S_b + S_4,
\end{equation}
must be tackled. This interesting possibility is left for future work. In the following, we only consider the dynamics driven by the brane action $S_b$.

\subsection{Tolman-Bondi Space-Times}
Tolman-Bondi space-times in extra-dimensions have been extensively studied \cite{Sil:1994ry,Banerjee:1994us,Ghosh:2001fb,Debnath:2002zz,Ghosh:2006ab,Quintavalle:2008,Mena:2009ap}. Here we review some of their salient properties.
We consider a $(d+4)$-dimensional universe filled with a pressureless fluid and a spatially dependent energy density $\rho(r,t)$ whose metric is given by the
extension of the Tolman-Bondi solution to $(d+4)$-dimensions
\begin{equation}
ds^2 =-dt^2 + e^{\lambda(r,t)} dr^2 + R^2(r,t) d\Omega^2_{d+2},
\end{equation}
where $d\Omega^2_{d+2}$ is the metric on the $(d+2)$-sphere, i.e there are  $n=d+2$ angular coordinates. The Einstein equations yield
\begin{equation}
\label{lambdadef}
e^{\lambda}= \frac{{R'}^2}{1+f(r)},
\end{equation}
where $'=d/dr$. The time-dependence of the metric  is specified by the Friedmann equation
\begin{equation}
\label{friedmann}
\dot R^2 = \frac{M(r)}{R^{n-1}} + f(r),
\end{equation}
where $\dot ~=d/dt$ and
where $f(r) $ is an arbitrary dimensionless function of the radius $r$.
When $f(r) = 0$, the Tolman-Bondi Universe reduces to a Friedmann-Lemaitre model with no inhomogeneity if we
assume a cosmically simultaneous Big-Bang time, i.e. $t_0=t_0(r)=$constant independently of $r$. The Big-Bang time appears as an integration function $t_0=t_0(r)$ so that each shell has its own big-bang time, a feature which differs from what happens in the Friedmann-Lemaitre  limit unless $t_0$ is explicitly chosen to be $r$-independent.
In this case we have $R(r,t)= r a(t)$ where $a(t)$ is the scale factor of the Friedmann-Lemaitre Universe.

The mass function $M(r)$ is defined by
\begin{equation}
\rho= \frac{n M'}{2\kappa_{4+d}^2 R^n R'},
\end{equation}
or equivalently
\begin{equation}
M(r)= \frac{2\kappa_{4+d}^2}{n }\int_0^r dr \rho R^n R' ,
\end{equation}
where $\rho(r,t)$ has dimension $(d+4)$.
In the flat Friedmann-Lemaitre case, we have $\rho(r,t)= \rho(t)\propto a^{-(n+1)}$ and $M(r)\propto r^{n+1}$.
We can integrate the Friedmann equation to obtain
\begin{equation}
R(t,r) = r \left(1 + \frac{n+1}{2} \sqrt{\frac{M}{r^{n+1}} }(t-t_0)\right)^{2/(n+1)},
\end{equation}
such that at $t = t_0$, $R(t_0) = r$. Choosing the time origin to be  $t_0 = 2 \sqrt{r^{n+1}/M}/(n+1)$, we can write
\begin{equation}
R(t,r)  = r \left(\frac{t}{t_0}\right)^{2/(n+1)} = \left(\frac{n+1}{2}\right)^{2/(n+1)} M^{1/(n+1)} t^{2/(n+1)}.
\end{equation}
We then obtain
\begin{equation}
e^{\lambda}= \left(\frac{t}{t_0}\right)^{4/(n+1)},
\end{equation}
which specifies the homogeneous Friedmann-Lemaitre models.

When  $f(r) \neq 0$, we can follow  the approach of Debnath and Chakraborty \cite{Debnath:2002fd} where the inverse of the Friedmann equation (\ref{friedmann}) is used to obtain $t= t(R,r)$
\begin{eqnarray}
\label{solt2}
t-t_0 = \frac{2}{n+1}  M^{-1/2} r^{(n+1)/2} ~_2F_1\left[\frac{1}{2},\frac{1+n}{2(n-1)},\frac{3n-1}{2(n-1)}, - \frac{R^{n-1}f}{M}\right] \nonumber \\
- \frac{2}{n+1}  M^{-1/2} R^{(n+1)/2} ~_2F_1\left[\frac{1}{2},\frac{1+n}{2(n-1)},\frac{3n-1}{2(n-1)}, - \frac{R^{n-1}f}{M}\right],
\end{eqnarray}
where at $t = t_0$, $R(t_0) = r$.
We can now simply use $R' = (\partial t/\partial r)/(\partial t/\partial R)$ to obtain the spatial derivative of $R$
\begin{eqnarray}
R'(t,r) &=& \pm (1+n) R \left(\frac{f'}{f}-\frac{M'}{M}\right) \mp ~_2F_1\left[1,\frac{n}{n-1},
\frac{3n-1}{2(n-1)},-\frac{R^{n-1}f}{M}\right] \times  \nonumber \\
&&  \frac{R^n}{M}\left(f+\frac{M}{R^n-1}\right) \left[(1+n)\frac{f'}{f}-2 \frac{M'}{M}\right] .
\end{eqnarray}
from which we can obtain $\lambda$.
When  $n=3$ and therefore $d=1$, (\ref{solt2}) simplifies considerably and we obtain
\begin{equation}
R^2 = (t-t_0)^2 f \pm 2(t-t_0) \sqrt{M} ,
\end{equation}
which gives an explicit description of the five-dimensional geometry. When $f=0$, we find that $R^2(r,t)= 2(t-t_0)\sqrt{M}$ corresponding
to a scale factor growing in $(t-t_0)^{1/2}$.

In the following, we will consider models where $f(r)$ vanishes at the origin of space $r=0$ in such a way that
$f(r)\ll M(r)/R^{n-1}$ close to the origin. As $M(r) \propto r^{n+1}$ there, this implies that $f(r)$ must go to zero faster than
$r^2$ locally. This guarantees that the curvature vanishes locally in a neighbourhood of the origin and the extra-dimensional geometry is locally the one of a flat Friedmann-Lemaitre model. In particular, this choice specifies that the origin, on top of being the centre of the distribution of matter, is  a special point of space where the geometry is nearly homogeneous and isotropic. The geometry at large far away from the origin remains isotropic but deviates from homogeneity as soon as $f(r)$ becomes relevant in the Friedmann equation.

\section{Mirage Cosmology}

In this section we will construct the mirage cosmology set up, first describing the higher dimensional embedding of our 3-brane and second, by deriving the dynamical equations that rule the induced
4 dimensional metric.

\subsection{Brane embedding}

We consider a D3 brane embedded in the Tolman-Bondi background in $(d+4)$ dimensions that we have presented in the last section.
The extra dimensional matter distribution is comoving with the energy-momentum tensor
\be
T^{AB}=\rho(r,t) u^A u^B ,
\ee
where matter particles are comoving in the bulk and spherically distributed with $u^A= \delta^A_0$.
In order to preserve spherical symmetry of the induced matter distribution on the brane with the induced  energy momentum tensor
\be
T_{\mu\nu}= T_{AB} \frac{\partial X^A}{\partial x^\mu}\frac{\partial X^B}{\partial x^\nu} ,
\ee
where $X^A(x^\mu)$ is the embedding of the brane with coordinates $x^\mu$ in the Tolman-Bondi Universe with coordinates $X^A$,
we choose to embed the brane as 
\begin{equation}
 x^0= t, \hspace{2cm}  x^1= r,
\end{equation}
where $x^0$ and $x^1$ are the brane coordinates.
Matter is then spherically symmetric on the brane with
\be
T_{\mu\nu}= \rho(x^0,x^1) u_\mu u_\nu,
\ee
where
\be
u_\mu= u_A \frac{\partial X^A}{\partial x^\mu},
\ee
is explicitly $u_\mu= -\delta_\mu^0$ as $u_A=- \delta^0_A$ and
$\partial X^0/\partial x^\mu = \delta^0_\mu$.

\begin{figure}
\begin{center}
\includegraphics[width=8cm]{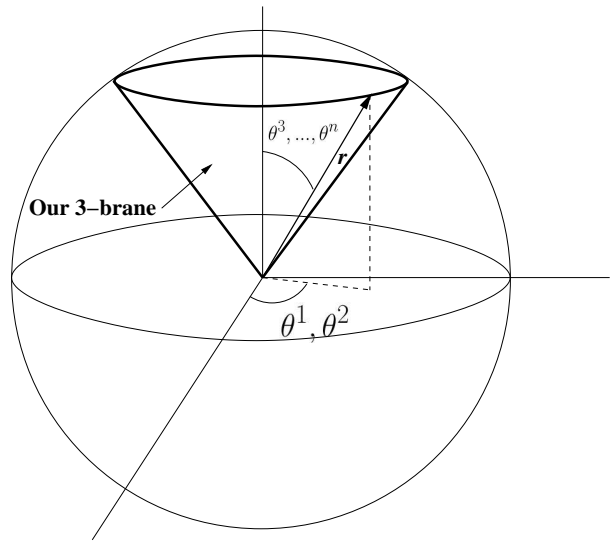}
\caption{\label{scheme} Schematic description of our higher dimensional embedding. Our 3-brane is represented by the cone defined by the fixed angles $\theta^3,...,\theta^n$ and the freely rotating $\theta^1$ and $\theta^2$.}
\end{center}
\end{figure}

This choice of embedding implies that the 3-brane corresponds to a two dimensional cross-section of the $(d+2)$-sphere.
Using the $(d+2)$-rotational invariance, we can always rotate the cross-section to align it with a planar section of the $(d+2)$ sphere parameterised by the angles
$\theta^1$ and $\theta^2$. Figure \ref{scheme} illustrates this embedding. Our 3-brane consists of the cone defined by the angles
$\theta^3,...,\theta^n =$ constant and the free angles $\theta^1$ and $\theta^2$.

Denoting by $\chi$ and $\phi$ the angles of the 2-sphere on the world-volume of the brane at constant $x^1$, we find that the embedding is
characterised by the two fields
\begin{equation}
\theta^1= \theta^1(t, r,  \phi, \chi), \hspace{1cm}  \theta^2= \theta^2( t, r, \phi, \chi),
\end{equation}
The other coordinates on the $(d+2)$-sphere are in the normal space to the brane and such that
\be
\theta^i=\theta^i_0, \hspace{2cm}  i\ge 3,
\ee
where $\theta^i_0$ are constant angles.

We are interested in the  geometry seen by a comoving observer on the brane characterised by the velocity vector
$v^\mu= \delta^\mu_0$ which differs from the velocity vector of matter particles as seen from the brane $u^\mu= \tilde g^{tt} \delta^\mu_0$
when $\theta^{1,2}$ are not time independent.
The spatial geometry seen by such a comoving observer is defined by the metric
\be
h^{\mu\nu}= \tilde g^{\mu\nu}+ v^{\mu}v^\nu,
\ee
satisfying $h_{\mu\nu}v^\mu=0$. On the space orthogonal to $v^\mu$ it reduces to $\tilde g^{ij}$ where $x^{i,j}=(r,\chi,\phi)$. This metric is both inhomogeneous and anisotropic.
Moreover proper time for a comoving observer at rest on the brane
\be
d\tau^2= - \tilde g_{tt} dt^2,
\ee
is also defined inhomogeneously in space. This is due to the $(t,r)$ dependence of $\theta^{1,2}$.

In the following, we will concentrate on the vicinity of the origin of space where we have assumed that the curvature effects in the bulk can be neglected. In the neighbourhood of this point, the bulk Tolman-Bondi metric reduces to a Friedmann-Lemaitre space-time. On the other hand, on the brane,  the local geometry defined by the induced metric is neither homogeneous nor isotropic there. An observer analysing the geometry of its local patch would see the Universe as both inhomogeneous and anisotropic. This is the case despite the fact that matter as seen on the brane is spherically symmetrically distributed according to $\rho(r,t)$.  We will study the long term dynamics of the geometry around the origin and see when the local geometry  isotropises and becomes homogeneous up to points where  the curvature effects in the bulk measured by $f(r)$ cannot be neglected anymore.

The dynamics of the brane, solely governed by its world volume action and depending on the tension $T_3$, depends on the induced metric.
With the chosen embedding, we have for the induced metric:
\begin{eqnarray}
ds^2 &=&    \tilde g_{tt} d  t^2 +   \tilde g_{rr} d  r^2 + 2 \tilde  g_{tr} d  t d  r +  \tilde g_{\phi\phi} d  \phi^2 + \nonumber \\
&~&  \tilde g_{\chi\chi} d  \chi^2 + 2   \tilde g_{\phi\chi}d\phi d \chi,
\end{eqnarray}
which can be related to the $(4+d)$ dimensional metric $G_{AB}$ using (\ref{inducedmetric}).
This gives explicitly
\begin{equation}
 \tilde  g_{rr}= e^{\lambda} + R^2 g_{ab} \frac{\partial  \theta^a}{\partial   r} \frac{\partial \theta^b}{\partial   r} =
e^{\lambda} + R^2 \left(\frac{\partial  \theta^1}{\partial   r}\right)^2 + R^2 \sin^2\theta^1\left(\frac{\partial \theta^2}{\partial   r}\right)^2,
\end{equation}
where $g_{ab}$ is the metric on the $(d+2)$-sphere and the $\theta^a$'s are spherical coordinates.
Here  we have  the  diagonal metric
with $g_{11}=1$, $g_{12}=0$ and $g_{22}=\sin^2 \theta^1$.
Similarly we have
\begin{eqnarray}
  \tilde g_{tt}= -1 + R^2  g_{ab} \frac{\partial  \theta^a}{\partial   t} \frac{\partial \theta^b}{\partial   t} =
-1 + R^2 \left(\frac{\partial  \theta^1}{\partial   t}\right)^2 + R^2 \sin^2\theta^1\left(\frac{\partial \theta^2}{\partial   t}\right)^2,  \\
  \tilde g_{tr}=  R^2 g_{ab} \frac{\partial   \theta^a}{\partial t} \frac{\partial   \theta^b}{\partial r} =
R^2 \frac{\partial  \theta^1}{\partial   t}\frac{\partial  \theta^1}{\partial   r} +
R^2 \sin^2\theta^1 \frac{\partial \theta^2}{\partial   t}\frac{\partial \theta^2}{\partial   r},
\end{eqnarray}
and
\begin{eqnarray}
  \tilde g_{\phi\phi}= R^2 g_{ab} \frac{\partial \theta^a}{\partial   \phi} \frac{\partial \theta^b}{\partial   \phi},
= R^2 \left(\frac{\partial  \theta^1}{\partial   \phi}\right)^2 + R^2 \sin^2\theta^1\left(\frac{\partial \theta^2}{\partial   \phi}\right)^2,
\\
\tilde  g_{\chi\chi}= R^2 g_{ab} \frac{\partial   \theta^a}{\partial  \chi} \frac{\partial   \theta^b}{\partial \chi}
= R^2 \left(\frac{\partial  \theta^1}{\partial   \chi}\right)^2 + R^2 \sin^2\theta^1\left(\frac{\partial \theta^2}{\partial   \chi}\right)^2
,\\
\tilde  g_{\phi\chi}= R^2 g_{ab} \frac{\partial   \theta^A}{\partial \phi} \frac{\partial   \theta^B}{\partial \chi}
=
R^2 \frac{\partial  \theta^1}{\partial   \phi}\frac{\partial  \theta^1}{\partial   \chi} +
R^2 \sin^2\theta^1 \frac{\partial \theta^2}{\partial   \phi}\frac{\partial \theta^2}{\partial   \chi}.
\end{eqnarray}
With these expressions, we can express the volume element of the brane in terms of the fields $\theta^{1,2}$. The dynamics of the brane is simply obtained
by looking at extremal world-volumes.

\subsection{Mirage dynamics}

We are now ready to write down the 3-brane world-volume action
(\ref{3braneaction}), find the corresponding Lagrangian in terms of  the higher dimensional angular coordinates and compute their equations of motion.
The action of the brane is given by the integral of the volume element
\begin{equation}
\sqrt{-\tilde  g} = \left[ - (\tilde g_{tt}\tilde  g_{rr} -\tilde  g_{tr}^2)(\tilde g_{\phi\phi}  \tilde g_{\chi\chi} -  \tilde g_{\phi\chi}^2)\right]^{1/2},
\end{equation}
leading to a Lagrangian
\begin{equation}
{\cal L}= -T_3 \left[ - (\tilde g_{tt}\tilde  g_{rr} -\tilde  g_{tr}^2)(\tilde g_{\phi\phi}  \tilde g_{\chi\chi} -   \tilde g_{\phi\chi}^2)\right]^{1/2}.
\end{equation}
We expand this Lagrangian to second order in the time and radial derivatives as we are interested in low energy modes. Hence we  get the Lagrangian
\begin{eqnarray}
{\cal L}&=& -T_3 R^2 e^{\lambda/2}\vert \sin \theta^1\vert \left\{1 + \frac{1}{2} R^2 e^{-\lambda} \left(\frac{\partial \theta^1}{\partial   r}\right)^2 + \frac{1}{2} R^2 e^{-\lambda} \sin^2 \theta^1 \left(\frac{\partial \theta^2}{\partial   r}\right)^2
\right. \nonumber \\
&~& \left. -\frac{1}{2} R^2 \left(\frac{\partial \theta^1}{\partial   t}\right)^2 -  \frac{1}{2} R^2  \sin^2 \theta^1 \left(\frac{\partial \theta^2}{\partial   t}\right)^2 \right\}  
\left( \frac{\partial \theta^1}{\partial   \phi}\frac{\partial \theta^2}{\partial   \chi}-
\frac{\partial \theta^1}{\partial   \chi}\frac{\partial \theta^2}{\partial   \phi}\right),
\end{eqnarray}
which can be conveniently cast in the form
\begin{equation}
{\cal L} = - J K L,
\end{equation}
where $J$ captures the angular properties of the embedding
\begin{equation}
\label{jeq1}
J \equiv
\left( \frac{\partial \theta^1}{\partial   \phi}\frac{\partial \theta^2}{\partial   \chi}-
\frac{\partial \theta^1}{\partial   \chi}\frac{\partial \theta^2}{\partial   \phi}\right),
\end{equation}
while $K/T_3$ is the volume element of an isotropic and homogeneous Friedmann-Lemaitre Universe in 4 dimensions
\begin{equation}
K \equiv T_3 R^2 e^{\lambda/2}s \sin \theta^1,
\end{equation}
where $s = |\sin \theta^1|/\sin \theta^1 = \pm 1$, and finally we have a reduced Lagrangian $L$ up to second order in the time and radial derivatives which
encompasses the dynamical properties of the 3-brane
\begin{eqnarray}
\label{defL}
L&\equiv& 1 + \frac{1}{2} R^2 e^{-\lambda} \left(\frac{\partial \theta^1}{\partial   r}\right)^2 + \frac{1}{2} R^2 e^{-\lambda} \sin^2 \theta^1 \left(\frac{\partial \theta^2}{\partial   r}\right)^2
 \nonumber \\
&~&  -\frac{1}{2} R^2 \left(\frac{\partial \theta^1}{\partial   t}\right)^2 -  \frac{1}{2} R^2  \sin^2 \theta^1 \left(\frac{\partial \theta^2}{\partial   t}\right)^2.
\end{eqnarray}
The dynamics of the brane are governed by the Euler-Lagrange equation for $\theta^1$ and $\theta^2$.
Noticing that the Lagrangian does not depend on $\theta^2$, we find a conservation equation
\begin{equation}
\partial_\mu A^\mu=0,
\end{equation}
where
\begin{equation}
A^\mu = \frac{\partial {\cal L}}{\partial (\partial_\mu \theta^2)},
\end{equation}
is the conjugate vector to $\theta^2$. We will see that this vector is not automatically  a constant, it has an explicit space-time dependence generically. Moreover, different choices for $A^\mu$ would lead to different solutions to the brane dynamics. In the following, we will see that the dynamical equations can be solved
for a subclass of vectors $A^\mu$. The general case is certainly interesting although challenging technically. Our analysis is general in this section.
Explicitly,  we have
\begin{eqnarray}
\label{defat}
A^t &=& J K R^2 \sin^2 \theta^1 \frac{\partial \theta^2}{\partial   t}, \\
\label{defar}
A^r &=&- JK R^2 e^{-\lambda} \sin^2 \theta^1 \frac{\partial \theta^2}{\partial   r}, \\
\label{defaphi}
A^\phi &=& KL \frac{\partial \theta^1}{\partial   \chi}, \\
\label{defachi}
A^\chi &=& -KL \frac{\partial \theta^1}{\partial   \phi}.
\end{eqnarray}
Equation (\ref{jeq1}) can now be written more compactly using (\ref{defaphi}) and (\ref{defachi}),
\begin{equation}
J = -\frac{1}{KL} \left(A^\chi \theta^2_{,\chi} + A^\phi \theta^2_{,\phi}\right).
\end{equation}
and then (\ref{defat}) and (\ref{defar}) can be recast as
\begin{eqnarray}
\label{dottheta2a}
\dot\theta^2 &=& -\frac{L A^t}{ R^2 \sin^2\theta^1(A^\chi \theta^2_{,\chi} + A^\phi \theta^2_{,\phi}) }, \\
\label{theta2r}
\theta^2_{,r} &=&  \frac{L A^r}{R^2 e^{-\lambda} \sin^2\theta^1 (A^\chi \theta^2_{,\chi} + A^\phi \theta^2_{,\phi})},
\end{eqnarray}
which upon substitution into Eq.~(\ref{defL}) give a quadratic equation for $L$ whose solutions are
\begin{equation}
L = (1 \pm \sqrt{1-4AC})/2A,
\end{equation}
where we have conveniently defined
\begin{eqnarray}
A &=& \frac{e^\lambda (A^r)^2 - (A^t)^2}{2 R^2 \sin^2\theta^1 (A^\chi \theta^2_{,\chi} + A^\phi \theta^2_{,\phi})^2 }, \\
C &=& 1 + \frac{1}{2} R^2 \left[ e^{-\lambda} {(\theta^1_{,r}})^2 - (\dot\theta^1)^2 \right].
\end{eqnarray}
In the following, we will retain the $+$ solution and impose that $|AC| \ll 1$ is such a way that $ L\sim 1/A$. The Euler-Lagrange equation for $\theta^1$ is a very complex partial derivative equation which does not seem to be tractable in full generality, however, by performing suitable simplifications we will find an elegant expression capable of being studied.

\subsection{Simplified dynamics}

We will reduce the dynamics of the brane in a Tolman-Bondi bulk by using simplifying assumptions. The first step consists in choosing a subclass of vectors $A^\mu$ for which the dynamics can be solved in a non-trivial way.
 So we will set \be A^r = A^\chi = 0\ee which immediately implies that   \be \theta^2_{,r} = \theta^1_{,\phi} = 0,\ee respectively and \be A = -(A^t)^2/2R^2 \sin^2\theta^1 (A^\phi)^2 (\theta^2_{,\phi})^2.\ee
In addition, we will assume that
$e^{-\lambda/2} \theta^1_{,r} \ll \dot{\theta^1}$. In effect, we consider that the dynamics is very close to being homogeneous. These choices will allow us to simplify the equation of motion for $\theta^1$, which can now take the form
\begin{equation}
\label{theequationofmotion}
\ddot\theta^1 +  \left( 4 \frac{\dot R}{R} + \frac{1}{2}\dot\lambda + 2 \frac{\dot\theta^1_{,\chi}}{\theta^1_{,\chi}} + \frac{\dot\theta^2_{,\phi}}{\theta^2_{,\phi}} \right) \dot\theta^1 +  \frac{\cos \theta^1}{\sin \theta^1} (\dot\theta^1)^2 = \dot\theta^2 \sin^2\theta^1 \frac{\theta^2_{,\phi}}{\theta^1_{,\chi}} \frac{\partial}{\partial t}
\left( \frac{\theta^2_{,\chi}}{\theta^2_{,\phi}}  \right).
\end{equation}
Notice that the radial partial derivatives are not present.
When $|AC| \ll 1$ we find that $\theta^1_{,\chi}$ can be approximated by
\begin{equation}
\label{theta1chi}
\theta^1_{,\chi} \approx -\frac{(A^t)^2}{2 T_3 A^\phi e^{\lambda/2} R^4 s \sin^3\theta^1  (\theta^2_{,\phi})^2} .
\end{equation}
In the following we will assume that $A^t$ is constant in time.
Differentiating $\theta^1_{,\chi}$ with respect to time
we find that
\begin{equation}
\label{dottheta1chi}
\dot\theta^1_{,\chi} = -\theta^1_{,\chi} \left(4 \frac{\dot R}{R} + \frac{\dot\lambda}{2} +  3 \frac{\cos \theta^1}{\sin \theta^1} \dot\theta^1+
2 \frac{\dot\theta^2_{,\phi}}{\theta^2_{,\phi}} + \frac{\dot{A}^\phi}{A^\phi}\right),
\end{equation}
which upon substitution in (\ref{theequationofmotion}) gives
\begin{equation}
\ddot\theta^1 -  \left( 4 \frac{\dot R}{R} + \frac{1}{2}\dot\lambda +
 3\frac{\dot\theta^2_{,\phi}}{\theta^2_{,\phi}} + 2 \frac{\dot{A}^\phi}{A^\phi} \right) \dot\theta^1- 5 \frac{\cos \theta^1}{\sin \theta^1} (\dot \theta^1)^2  =  \dot\theta^2 \sin^2\theta^1 \frac{\theta^2_{,\phi}}{\theta^1_{,\chi}} \frac{\partial}{\partial t}
\left( \frac{\theta^2_{,\chi}}{\theta^2_{,\phi}}  \right).
\end{equation}
This is the partial differential equation which governs the brane dynamics.

We have seen in Eq.~(\ref{dottheta2a}) that $\dot\theta^2 = - L A^t /A^\phi \theta^2_{,\phi} R^2 \sin^2 \theta^1$, which under our approximations gives
\begin{equation}
\label{dottheta2}
\dot\theta^2 \approx 2 \frac{A^\phi}{A^t} \theta^2_{,\phi}.
\end{equation}
As we have required  $\theta^1_{,\phi} = 0$ we find using Eq.~(\ref{theta1chi}) that $\theta^2$ must be linear in $\phi$, i.e. $\theta^2_{,\phi} = c_0$ where $c_0$ is a constant and
\begin{equation}
\theta^2 = c_0 \phi +   2 c_0 \int_{t_0}^t \frac{A^\phi}{A^t} dt .
\end{equation}
Notice that $\theta^2_{,r}=0$ implies that $A^\phi/A^t$ must be $r$-independent.

We have now obtained the equations of motion for $\theta^1 = \theta^1(t,\chi)$ and the general form of the solutions for $\theta^2 = \theta^2(t,\phi,\chi)$. One must therefore solve
\begin{equation}
\ddot\theta^1 -  \left( 4 \frac{\dot R}{R} + \frac{1}{2}\dot\lambda +
  2 \frac{\dot{A}^\phi}{A^\phi} \right) \dot\theta^1 - 5 \frac{\cos \theta^1}{\sin \theta^1} (\dot \theta^1)^2  = \sin^2 \theta^1\, \frac{\dot\theta^2 \,     \dot\theta^2_{,\chi}}{\theta^1_{,\chi}}.
  \label{eq1}
\end{equation}
The source term depends on 
$A^\phi$ and $A^t$. We have therefore reduced the brane dynamics to a single differential equation.

\section{The Brane Geometry Close to the Origin}

In this section we will solve (\ref{eq1}) by decomposing $\theta^1$  and $A^\phi$ in their respective time and $\chi$-dependent parts. We will see that the solution has a late time attractor.

\subsection{Local Attractor}
Even the simplified dynamics are difficult to analyse.
We concentrate on the interesting case of a 3-brane embedding in a  Tolman-Bondi Universe where the curvature effects measured by $f(r)$ can be neglected in the neighbourhood of the origin. In this case, the extra-dimensional dynamics in a local patch becomes:
\begin{equation}
R(t,r)=r \left(\frac{t}{t_0}\right)^{2/(n+1)},\hspace{1cm} e^{\lambda}= \left(\frac{t}{t_0}\right)^{4/(n+1)},
\end{equation}
and we focus on the induced metric on the brane which takes the simplified form
\begin{eqnarray}
\tilde g_{tt} &=& -1 + R^2 (\dot\theta^1)^2 + R^2 \sin^2\theta^1 (\dot\theta^2)^2 , \\
\tilde g_{rr} &=& e^\lambda , \\
\tilde g_{\phi\phi} &=& R^2 \sin^2 \theta^1 (\theta_{,\phi}^2)^2 , \\
\tilde g_{\chi\chi} &=& R^2 (\theta^1_{,\chi})^2 +R^2 \sin^2 \theta^1 (\theta^2_{,\chi})^2  , \\
\tilde g_{\phi\chi}&=& R^2 \sin^2 \theta^1 \theta^2_{,\chi} \theta^2_{,\phi} .
\end{eqnarray}
We will analyse the geometry on the brane when we have the explicit solutions of the equations of motion.

We are interested in embeddings where the lapse function does not deviate too much from 1.
We consider then
\begin{equation}
\ddot\theta^1 -  \left( 4 \frac{\dot R}{R} + \frac{1}{2}\dot\lambda +
  2 \frac{\dot{A}^\phi}{A^\phi} \right) \dot\theta^1 - 5 \frac{\cos \theta^1}{\sin \theta^1} (\dot \theta^1)^2 = I.
\end{equation}
where the forcing term is
\begin{equation}
I=\sin^2 \theta^1 \, \frac{\dot\theta^2 \,    \dot\theta^2_{,\chi}}{\theta^1_{,\chi}}.
\end{equation}
We look for  a solution for $\theta^1$ locally such that  $\theta^1 \approx \theta_0$ and we write
\begin{equation}
\theta^1 = \theta_0 -\frac{(A^t)^2}{2T_3 s \sin^3 \theta_0 M_0^6 c_0^2 r^4  g e^{\lambda/2} R^4}\int \frac{1}{f} \, d\chi  + \alpha(t) ,
\end{equation}
where $\theta_0$ is a fixed angle arising as an integrating constant, $\alpha (t)$ an unknown function yet to be determined
and we have introduced
\begin{equation}
A^\phi(r, t,\chi) = M_0^6 r^4 f(\chi) g(t),
\end{equation}
in order to separate the time and $\chi$-dependent contributions that make up $A^\phi$. Here, $M_0$ is an unspecified energy scale.
Notice that the $r$-dependence cancels if and only if $A^t$ varies in $r^4$ as required by $\theta^2_{,r}=0$.
We now consider that $\chi $ measures the small deviations from $\theta_0$ and approximate $f = 1+ f_1 \chi$,
\begin{equation}
\theta^1 = \theta_0 -\frac{(A^t)^2}{2T_3M^6_0 c_0^2 s \sin^3 \theta_0 r^4 g e^{\lambda/2} R^4} \chi  + \alpha(t).
\end{equation}
We  identify the constant of motion $A^t$ as
\begin{equation}
(A^t)^2 = 2 T_3 M^6_0 c_0^2 s \sin^3 \theta_0 r^8 g_0 ,
\end{equation}
and  write the solution for $\theta^1$ as
\begin{equation}
\theta^1 = \theta_0- \frac{g_0}{g} e^{-\lambda/2} \left( \frac{r}{R}\right)^4 \chi  + \alpha(t) ,
\end{equation}
which is $r$-independent, hence satisfying our assumption on the small $r$ variations of $\theta^1$.
At $t=t_0$ we have that
\begin{eqnarray}
\theta^1 &=& \theta_0 - \chi, \\
\theta^2 &=& c_0 \phi ,
\end{eqnarray}
where we have chosen $\alpha (t_0)=0$. Thus initially, the spatial part of the induced metric is
\begin{eqnarray}
\tilde g_{rr} &=& 1, \\
\tilde g_{\phi\phi} &=& r^2 \sin^2 \theta^1, \\
\tilde g_{\chi\chi} &=& r^2, \\
\tilde g_{\phi\chi} &=& 0,
\end{eqnarray}
which corresponds to flat space. As time passes, the metric is deformed and deviates from flat space.
On the other hand the lapse function $\tilde g_{tt}$ can not be uniformly equal to $-1$  even initially, implying that time flows differently at different points
of space.

The analysis of the equation of motion for $\theta^1$ is greatly simplified if one studies the evolution of small deviations from $\theta_0= \pi/2$, i.e. we consider an expansion in small $\chi$ around $\theta^1= \pi/2$. The general case around a general $\theta_0$ follows easily as we will see soon.
Substituting the previous ansatz for $\theta^1$ in the equation of motion and separating  the terms in $\chi$ from  the ones of order zero in $\chi$, we obtain two second order differential equations for $\alpha$ and $g$
\begin{equation}
\ddot\alpha - \left( 4 \frac{\dot R}{R} + \frac{1}{2}\dot\lambda +
  2 \frac{\dot g}{g} \right) \dot\alpha  = I_\alpha ,
\end{equation}
where
$I_\alpha=- (2 f_1 g_0 M_0^6/T_3) e^{\lambda/2} (g/g_0)^3 (R/r)^4$
and
\begin{eqnarray}
\label{gequation}
&&\frac{\ddot g}{g}  - \left(\frac{\dot g}{g}\right)^2 + 4\frac{\ddot R}{R} - 4 \left(\frac{\dot R}{R}\right)^2 + \frac{1}{2} \ddot \lambda - \left( \frac{\dot g}{g} + 4 \frac{\dot R}{R} + \frac{1}{2} \dot \lambda \right)^2  - \nonumber \\
&&\left(2\frac{\dot g}{g} + 4 \frac{\dot R}{R} + \frac{1}{2} \dot \lambda \right)   \left(\frac{\dot g}{g} + 4 \frac{\dot R}{R} + \frac{1}{2} \dot \lambda \right) = I_g ,
\end{eqnarray}
where
$I_g=- (4 f_1^2g_0 M_0^6/T_3) e^{\lambda} (g/g_0)^4 (R/r)^8$.
The solutions to these equations can be easily obtained.

First of all it follows from the $\alpha$-equation that the homogeneous solution when the forcing term vanishes is
\begin{equation}
\label{alphasol}
\alpha(t) = \dot \alpha_0 \int_{t_0}^t dt'  e^{\lambda/2} \left(\frac{R}{r}\right)^4 \left(\frac{g}{g_0}\right)^2 .
\end{equation}
The $g$-equation (\ref{gequation}) is equivalent to a Riccati equation
\begin{equation}
\dot V= 3 R^{-4} e^{-\lambda/2} V^2 + R^4 e^{\lambda/2} I_g ,
\end{equation}
and we have defined
\begin{equation}
V= R^4 e^{\lambda/2} \left(\frac{\dot g}{g} + 4 \frac{\dot R}{R} + \frac{\dot \lambda}{2}\right) ,
\end{equation}
Putting $b= 3 R^{-4} e^{-\lambda/2}$ and $c=R^4 e^{\lambda/2} I_g$, the Riccati equation becomes a Sturm-Liouville equation
\begin{equation}
\ddot X + \left[ bc -b^{1/2} \frac{d^2}{d t^2}\left(\frac{1}{b^{1/2}}\right) \right] X=0,
\end{equation}
where the following correspondence
\begin{equation}
V=-\frac{\dot \omega}{b w}, \hspace{1cm} X=b^{-1/2} \omega,
\end{equation}
applies.
The homogeneous solution ($c=0$)  is
\begin{equation}
X=b^{-1/2}\left(c_1 + c_2 \int_{t_0}^t dt' b(t')\right) ,
\end{equation}
where $c_1$ and $c_2$ are integration constants.
As a result, we find that the solution of the homogeneous $g$-equation is
\begin{equation}
\label{local}
g = g_0 \left(\frac{t_0}{t}\right)^{10/(n+1)} \left| 1 + c_3 \left[ \left(\frac{t_0}{t}\right)^{(9-n)/(n+1)}-1\right] \right|^{-1/3} ,
\end{equation}
where $c_3$ is a constant.
 We find that as long as the term within the power $-1/3$ does not vanish, a statement
which only depends on the initial conditions, the asymptotic behaviour of the homogeneous solution is
\begin{equation}
g \approx  \gamma_g t^{-10/(n+1)} ,
\label{gat}
\end{equation}
where $\gamma_g$ is a constant. Using this result and substituting in (\ref{alphasol}) one finds that the asymptotic behaviour of $\alpha$ is
\begin{equation}
\alpha \approx \gamma_\alpha t^{(n-9)/(n+1)} + \alpha_\infty ,
\label{al}
\end{equation}
where $\gamma_\alpha$ and $\alpha_\infty$ are constants, i.e., for $n<9$, $ \alpha$ converges to a constant at infinity.
As $I_g={\cal{O}}(t^{-20/(n+1)})$ falls off faster than $t^{-2}$ and $I_\alpha={\cal O}(t^{-20/(n+1)})$ faster than $\ddot\alpha={\cal{O}}(t^{-(n+11)/(n+1)})$ for $n<9$, we find that the
homogeneous solutions ($\ref{gat}$) and ($\ref{al}$) are late time attractors where the influence of the forcing term $I_g$ and $I_\alpha$ becomes negligible.
The same results hold for any $\theta_0$ as we can neglect the quadratic term in $\dot\theta^1$ in Eq.~(\ref{eq1}) as the solution of the homogeneous equation behaves as $\theta^1=O(\beta_1 +\beta_2 t^{(n-9)/(n+1)})$ while $I={\cal O}(t^{-20/(n+1)})$. Hence, in the neighbourhood of each point $\theta_0$, we find that $\theta^1$ and $\theta^2$ become time-independent.
As a result the metric converges to
\begin{equation}
\label{dsinfty}
ds^2_\infty = -dt^2 + e^\lambda dr^2 + R^2 \left[ \sin^2\theta^1 c_0^2 d\phi^2 + \left((\theta^1_{,\chi})^2_{t\rightarrow\infty} + \sin^2\theta^1 (\theta^2_{,\chi})^2_{t\rightarrow\infty} \right) d\chi^2 + c_0 \sin^2 \theta^1 (\theta^2_{,\chi})_{t\rightarrow\infty} d\phi d\chi \right],
\end{equation}
and on the attractor $\theta^2_{,\chi}$ evolves as
\begin{equation}
\theta^2_{,\chi} = \frac{n+1}{n-9} f_1 M_0^3 t_0 \left(\frac{2 g_0}{T_3}\right)^{1/2} \left[ \left(\frac{t_0}{t}\right)^{(9-n)/(n+1)}-1\right],
\end{equation}
which indeed tends  asymptotically to a constant value
$(\theta^2_{,\chi})_{t\rightarrow \infty}$.
Also $\theta^1_{,\chi}$ is equal to minus one  on the attractor.
As a result the mapping between $(\chi,\phi)$ and $(\theta^1,\theta^2)$ becomes in the late time regime
\begin{eqnarray}
\theta^1_{t \rightarrow\infty} &=& \theta_0 -\chi +\alpha_\infty ,\nonumber \\
\theta^2_{t \rightarrow \infty}&=& c_0\phi+ (\theta^2_{,\chi})_{t\rightarrow \infty} \chi .
\end{eqnarray}
This is a linear transformation with constant coefficients. The transformation of the metric $g_{ab}$ on the two sphere
\begin{equation}
g_{ij}=g_{ab}\frac{\partial\theta^a_{t\rightarrow\infty}}{\partial x^i}\frac{\partial \theta^b_{t\rightarrow\infty}}{\partial x^j}
\end{equation}
where $x^i=(\phi,\chi)$ is a linear transformation defined globally for any time. The asymptotic metric is then
\begin{equation}
ds^2_\infty = -dt^2 + e^\lambda dr^2 + R^2 g_{ij} dx^i dx^j,
\end{equation}
which is equivalent to
\begin{equation}
ds^2_\infty = -dt^2 + e^\lambda dr^2 + R^2 g_{ab} d\theta^a_{t\rightarrow \infty} d\theta^b_{t\rightarrow \infty}.
\end{equation}
Hence we find that the asymptotic metric is isotropic,
the induced space-time on the brane isotropises and converges to a Friedmann-Lemaitre space-time close to the origin, of the radiation dominated type for $n = 3$.

We will see in the next section that the solution obtained here locally around each point specified by $\theta_0$ in a small $\chi$ expansion is in fact a global solution of the equations of motion. The analysis of this section shows that the global solution we will find in the next subsection is an attractor.

\subsection{Global Attractor}

We have seen that the fields $\theta^1$ and $\theta^2$ become locally time-independent in any neighbourhood of $\theta_0$ and that this solution is an attractor, i.e. all solutions around $\theta_0$ converge towards a unique long time solution determined essentially by the asymptotic behaviour of the function $g(t)\sim t^{-10/(n+1)}$. Here we will find a global solution which corresponds to this attractor, i.e. we will find a solution for very large time which matches the local behaviour of the solution determined by $g(t)\sim t^{-10/(n+1)}$. In particular, we will find that the induced metric on this attractor is the one of a radiation dominated Friedmann-Lemaitre Universe when $n=3$.

The homogeneous solution of (\ref{eq1}) when the term in $(\dot \theta^1)^2$ is neglected is simply
\begin{equation}
\theta^1(t,r,\chi)= E(\chi) \int_{t_0}^t dt' R^4 e^{\lambda/2} (A^\phi)^2 + F(\chi).
\end{equation}
The first term is time dependent while the second one is only a function of $\chi$. This general solution must be compatible with the differential equation (\ref{theta1chi}) which can be obtained putting $E(\chi)=0$. In this case, the behaviour of $\theta^1 (\chi)$,
necessitates that  $(A^t)^2/A^\phi e^{\lambda/2} R^4 (\theta^2_{,\phi})^2$ be time and $r$ independent.
This is achieved by choosing $\theta^2_{,\phi}=c_0$  constant and $A^t$ time independent.
Moreover we choose $A^t \propto r^4$ and $A^\phi \propto r^4$ implying that $\theta^1_\chi$ is $r$-independent.
Similarly $A^\phi$ must be such that $A^\phi e^{\lambda /2}R^4$ is time-independent. This implies that
\begin{equation}
A^\phi \propto t^{-10/(n+1)} .
\end{equation}
Notice that the behaviours of $A^\phi$ and $A^t$ as a function of $r$ and $t$ match the asymptotic behaviour of the local solution as they are both proportional to $r^4$ and that $A^\phi$ has the same time dependence as the asymptotic behaviour of $g(t)$ around any background value $\theta_0$.
We can now integrate exactly (\ref{theta1chi}) and get
the global mapping between $\chi$ and $\theta^1$
\begin{equation}
\int d\theta^1 \vert \sin \theta^1\vert^3 =  -\frac{(A^t)^2}{2T_3 c_0^2 A^\phi e^{\lambda/2} R^4 } \chi,
\end{equation}
where the right hand side is independent of $r$ and $t$. Expanding the left hand side around a given $\theta_0$ to linear order gives the local solution of the previous section.

For this solution, the non-linear term in $(\dot \theta^1)^2$  of (\ref{theequationofmotion}) vanishes identically while the forcing term converges to zero $I={\cal O}(t^{-20/(n+1)})$.
Hence this global solution is valid at infinity and around each $\theta_0$  it is identical to the one obtained in the previous subsection from (\ref{local}) with $c_3=0$. Hence, we have found a global solution of the equation of motion valid in the late time regime and we have shown that this solution is an attractor as the solution (\ref{local}) converges to (\ref{gat}).

In conclusion, we have found that locally around the origin of space-time on the brane, the geometry isotropises in the long time regime. This implies that a local observer, comoving on the brane would eventually see a flat Friedmann-Lemaitre space-time in its local patch. This result is special to the origin where we have assumed that the bulk effects of the curvature of space-time can be neglected. In a sense, we have shown that for a brane on which the matter density induced from the bulk is spherically symmetric and where the effects of the bulk curvature are negligible, the brane geometry eventually becomes homogeneous and isotropic locally. Of course, this is not the case away from the origin implying that interesting inhomogeneous effects on the propagation of light on the brane for instance may be induced. The study of the brane geometry at large away from the origin is left for future work.

\subsection{Numerical Results}

In this section we will illustrate the previous results and show numerically that the attractor mechanism is at play for a wide range of initially anisotropic and inhomogeneous space-times.
First, we must demand that the late time solution is such that $\tilde g_{tt} \approx -1$, i.e., that $R^2 (\dot\theta^1)^2 \ll 1$ and
$R^2(\dot\theta^2)^2 \ll 1$. As
we have seen, $A^\phi = M^6r^4 g(t) f(\chi)$ and $(A^t)^2 \approx 2 T_3 M^6 c_0^2 r^8 g_0$ then we have from (\ref{dottheta2}) that
\begin{equation}
\label{cond1}
R^2 (\dot\theta^2)^2 = \frac{8 f^2 M_0^6}{T_3} g_0 r^2 \left(\frac{t_0}{t}\right)^{16/(n+1)} \ll 1.
\end{equation}
Second, we must also require $4|AC| \ll 1$ so that we can use the approximation $L \approx 1/A$ and therefore, we have the additional constraint
\begin{equation}
\label{cond2}
4 |AC| \approx \frac{4 T_3 c_0^2}{M_0^6 f^2} \frac{1}{r^2 g_0} \left(\frac{t}{t_0}\right)^{16/(n+1)} \ll 1.
\end{equation}

\begin{figure}
\begin{center}
\includegraphics[width=8cm]{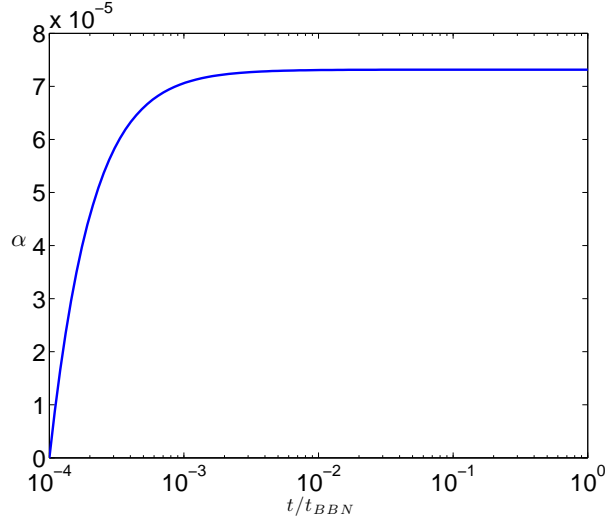}
\caption{\label{fig1} Evolution of $\alpha(t)$ with time for $T_3 = 1$GeV, $M_0 = 1$GeV, $n=3$, $f_1 = 10^{-15}$, $\chi = 0$, $g_0 = g(t=0) = 1.2 \times 10^{-41}$, $\dot g(t=0) = -9.5 g_0/(n+1)$, $c_0 = 4.4\times 10^{-8}$, $\alpha(t=0) = 0$, $\dot\alpha(t=0) = 10^{-4}$ and $r = 10^{-4} t_{BBN}$. One observes that $\alpha$ approaches a constant value at late times as we expected from  Eq.~(\ref{al}).}
\end{center}
\end{figure}

\begin{figure}
\begin{center}
\includegraphics[width=8cm]{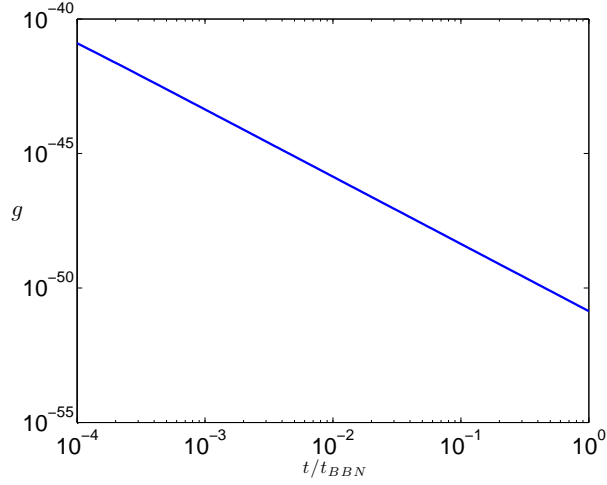}
\caption{\label{fig2} Evolution of $g(t)$ for the same parameters of Fig.~\ref{fig1}. We confirm that $g(t)$ decays as a power law given by Eq.~(\ref{gat}).}
\end{center}
\end{figure}

\begin{figure}
\begin{center}
\includegraphics[width=8cm]{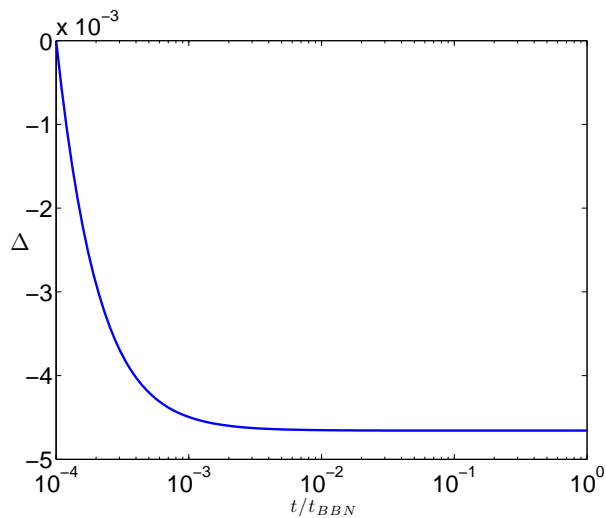}
\caption{\label{fig3} Evolution of $\Delta = 100 (\theta^1-\pi/2)/(\pi/2)$ for $\chi = 0$ which is the deviation of $\theta^1$ from $\theta_0 =\pi/2$ in percentage. This example clearly shows that the deviation is very small which ensures that our approximation $\sin \theta^1 \approx \sin \theta_0$ is valid during the whole time interval analysed.}
\end{center}
\end{figure}

\begin{figure}
\begin{center}
\includegraphics[width=8cm]{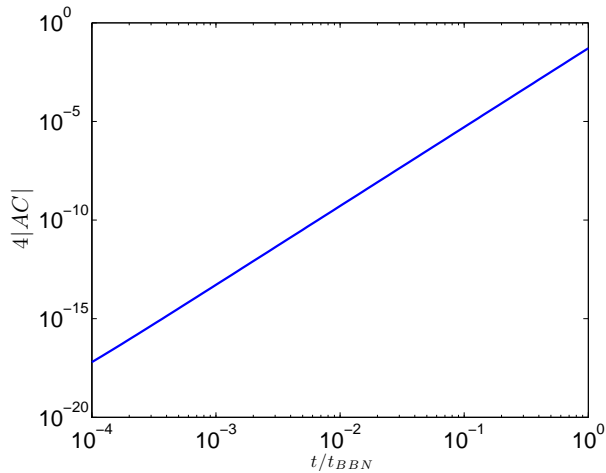}
\caption{\label{fig14} The time evolution of $4|AC|$ which we require to be smaller than unity for our simplification $L \approx 1/A$ to be valid. This is indeed the case for the time interval we are covering and for our choice of parameter values.}
\end{center}
\end{figure}

\begin{figure}
\begin{center}
\includegraphics[width=8cm]{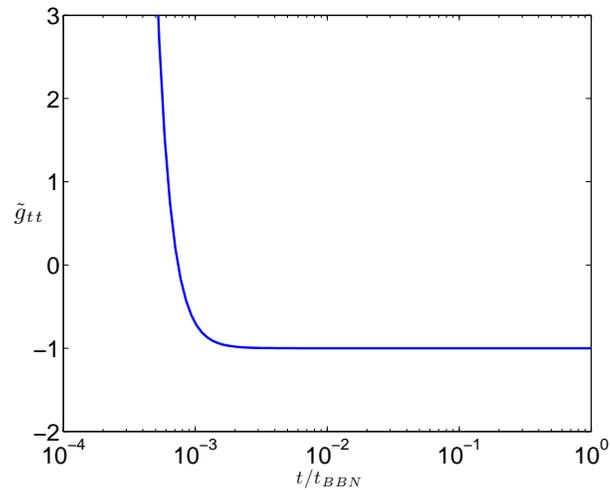}
\caption{\label{fig8} Evolution of the metric component $\tilde g_{tt}$.
To recover a Friedmann-Lemaitre metric this quantity must be $-1$ which indeed happens for our solution at late times.}
\end{center}
\end{figure}

\begin{figure}
\begin{center}
\includegraphics[width=8cm]{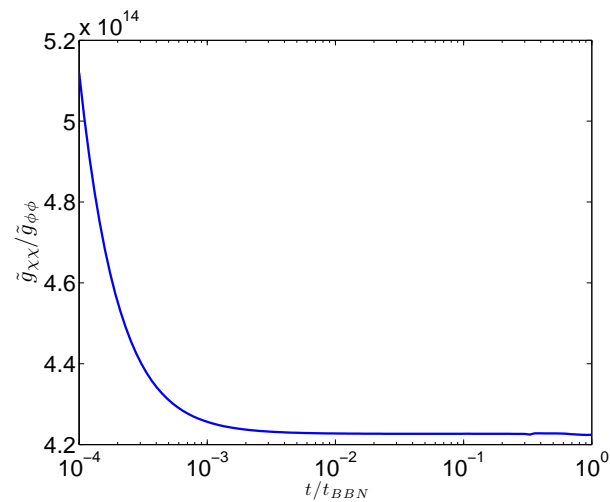}
\caption{\label{fig12} The time evolution of $\tilde g_{\chi\chi}/\tilde g_{\phi\phi}$. This ratio approaches a constant at late times as expected from Eq.~(\ref{dsinfty}).}
\end{center}
\end{figure}

\begin{figure}
\begin{center}
\includegraphics[width=8cm]{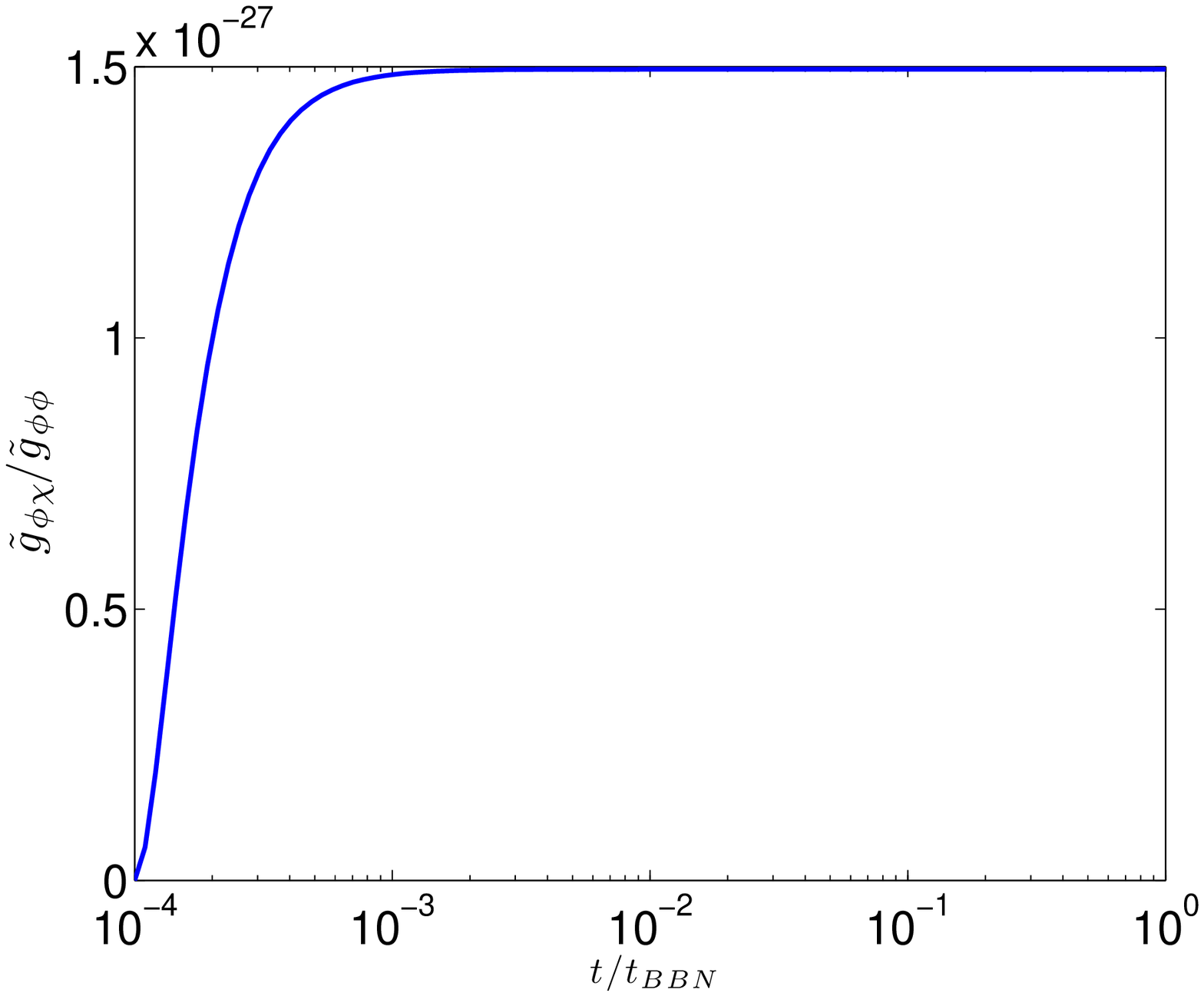}
\caption{\label{fig13} The time evolution of $\tilde g_{\phi\chi}/\tilde g_{\phi\phi}$ approaches a constant value at late times as we expected from Eq.(\ref{dsinfty}).}
\end{center}
\end{figure}

When numerically solving the equations of motion we have started the evolution at $t_0 = 10^{-4} t_{BBN}$ where the time of Big Bang Nucleosythesis is $t_{BBN} \approx 10^{26}$GeV$^{-1}$. We will compute solutions for the scale of the Hubble horizon at that time, i.e. $r = t_0$. We set $T_3 = 1$GeV, $M_0 = 1$GeV, $n=3$, $f_1 = 10^{-15}$, $\chi = 0$ and in order to satisfy conditions (\ref{cond1}) and (\ref{cond2}) we have  used
$g_0 = 1.2 \times 10^{-41}$ and $c_0 = 4.4\times 10^{-8}$.  The small size of $f_1$ is chosen to prevent the source terms $I_\alpha$ and $I_g$ to be of any importance.
As initial conditions we have used $\alpha(t=0) = 0$, $\dot\alpha(t=0) = 10^{-4}$, $g(t=0) = g_0$. If the evolution was already on the attractor we would have $\dot g = -10 g_0/(n+1)$. Instead we start slightly away from the attractor solution using  $\dot g(t=0) = -9.5 g_0/(n+1)$.
The result of the numerical solutions is shown in Figs.~\ref{fig1} to \ref{fig13}. In Fig.~\ref{fig1} we can observe the time dependence of $\alpha(t)$ and confirm that it quickly approaches a constant value $\alpha_\infty$ as we expect from Eq.~(\ref{al}).
On the other hand, it can be seen in Fig.~\ref{fig2} that $g(t)$ approaches the power law solution $g(t) \sim t^{-5/2}$ as we have seen in (\ref{gat}) for $n = 3$.

\noindent Figure \ref{fig3} shows that $\theta^1$ never deviates substantially from $\theta_0 = \pi/2$ which ensures that our approximations, i.e. neglecting the $(\dot\theta^1)^2$ term in Eq.~(\ref{eq1}) and making $\sin^3 \theta^1 \approx \sin^3 \theta_0$, are valid and under control.

\noindent We also demanded that $4|AC|\ll 1$ so that $L \approx 1/A$ which allows us to have a simple expression for $\theta^1_{,\chi}$ and be able to write Eq.~(\ref{theta1chi}). As we can see in Fig.~\ref{fig14} this is satisfied for our choice of parameters.

\noindent Figures \ref{fig8}, \ref{fig12} and \ref{fig13} show that indeed the metric on our 3-brane quickly isotropises at late times as  $\tilde g_{tt} \rightarrow -1$, $\tilde g_{\chi\chi}/\tilde g_{\phi\phi} \rightarrow$ constant,  $\tilde g_{\phi\chi}/\tilde g_{\phi\phi} \rightarrow$ constant
and $\tilde g_{\phi\chi}/\tilde g_{\phi\phi} $ too. This is what allows us to redefine the angular variables asymptotically and find that the spatial metric is the one of flat space.

\section{conclusion}

We have studied the dynamics of a 3-brane embedded in an inhomogeneous extra dimensional Universe. The mirage cosmology on the brane reveals
that the late time behaviour metric on the brane becomes isotropic and homogeneous in the vicinity of the origin,  where the effects of curvature in the bulk  can be neglected, even though the initial condition were not so. This isotropisation process is extremely interesting in view of the possible existence of extra dimension and the strongly motivated possibility that matter, even cosmological matter, could exist non-only in 4d but also in the extra dimensions. In this case, the dynamical interplay between the brane degrees of freedom and the extra dimensions becomes more complex than in the usual case of an extra dimensional symmetric space, e.g. anti de Sitter or Minkowski. Indeed, one may envisage situations where both matter would exist on the brane and in the extra dimensions, with possible exchanges between them. In this paper we have only analysed the mirage case where matter on the brane is subdominant and the dynamics are governed by the embedding of the brane in the time-dependent extra dimensional background. The coupling between matter on the brane and outside the brane  in an inhomogeneous context is left for future work.
The presence of inhomogeneities in the extra dimensions may even shed some light on thorny issues such as the acceleration of expansion of the 4d Universe or the role of dark matter. These tantalising possibilities are beyond the present work.

\begin{acknowledgments}
This work was carried out in the context of the Portugal-France Programa Pessoa bilateral agreement.
The work of NJN is supported by a Ci\^encia 2008 research contract funded by FCT.
JPM and NJN also acknowledge support from FCT through the projects PEst-OE/FIS/UI2751/2011, PTDC/FIS/102742/2008 and  grant CERN/FP/116398/2010. The authors are indebted to Daniele Steer for discussions and for comments on the manuscript.
\end{acknowledgments}

\end{document}